\documentclass{rspublic}
\usepackage{graphicx}
\usepackage{latexsym,amssymb}

\begin{document}

\title {Confinement induced instability of thin elastic film}

\author [A. Ghatak] {Animangsu Ghatak $^{\sharp}$ \footnotemark[1]}

\footnotetext[1]{To whom correspondence should be addressed; ({\em
aghatak@iitk.ac.in})}

\affiliation{$^{\sharp}$ Department of Chemical Engineering,
Indian Institute of Technology, Kanpur, UP 208016, India}

\label{firstpage}

\maketitle

\begin{abstract}{Confinement, Incompressibility}
A confined incompressible elastic film does not deform uniformly
when subjected to adhesive interfacial stresses but with
undulations which have a characteristic wavelength scaling
linearly with the thickness of the film. In the classical peel
geometry, undulations appear along the contact line below a
critical film thickness or below a critical curvature of the
plate. Perturbation analysis of the stress equilibrium equations
shows that for a critically confined film the total excess energy
indeed attains a minima for a finite amplitude of the
perturbations which grow with further increase in the confinement.
\end{abstract}

%\baselineskip = 2\baselineskip
\section{Introduction}
Spontaneous surface and interfacial instabilities of thin liquid
films have been studied in different contexts, e.g. the classical
Saffman-Taylor (Saffman \textit{et al} 1958) problem in a Hele
Shaw cell in which flow driven fingering patterns develop at the
moving interface of two viscous or viscoelastic liquids
(Homsy\textit{et al} 1981, Nittmann \textit{et al} 1985);
disjoining pressure induced rupturing and dewetting (Reiter
\textit{et al} 1992, Sharma \textit{et al} 1998) of ultra thin
viscous films; spiral instabilities in viscometric flow of a
viscoelastic liquid (Muller \textit{et al} 1989, McKinley
\textit{et al} 1995); and fingering instability and cavitation
during peeling a layer of viscoelastic adhesive (Fields \textit{et
al} 1976, Urhama \textit{et al} 1989). While most of these viscous
and viscoelastic systems have been well characterized
experimentally and theoretically, similar surface undulations of
confined thin elastic films pose a different kind of problem
despite geometric commonalities with many of the liquid systems.
The essential difference being that unlike in the liquid system
there is no flow of mass and consequent permanent deformation in
the elastic body, where the extent of deformation is governed by
the equilibrium of the external surface or body forces on the
material and the elastic forces developed.

The specific system that will be described in this paper is a thin
layer of elastic adhesive confined between a rigid and a flexible
plate. While the film remains strongly bonded to the rigid
substrate, the flexible plate is detached from it in the classical
peel geometry. High aspect ratio of such systems is noteworthy as
the lateral length scale far exceeds the thickness of the film
resulting in high degree of confinement for the adhesive. As a
result, adhesive stresses at the interface does not always result
in uniform deformation through out the whole area of contact,
rather spatially varying deformations (Gent \textit{et al} 1958,
Gent \textit{et al} 1969, Ghatak \textit{et al} 2000,
M$\ddot{o}$nch \textit{et al} 2001) attain lower energy for the
system. Experimentally we see the existence of a critical
thickness of the film or a critical curvature of the flexible
plate below which the contact line between the film and the
flexible plate does not remain straight, but turns undulatory with
a characteristic wavelength which increases linearly with the
thickness of the film (Ghatak \textit{et al} 2003). While
experimentally this phenomenon has been characterized, there is
not much understanding as what drives this instability in a
non-flow purely elastic system and how does the curvature of the
plate or the thickness of the film result in critical confinement
of the film. Here I present a perturbation analysis which
addresses these questions highlighting the duel effects of the
incompressibility of the elastic film and its confinement.
\section{Problem Formulation}
\begin{figure}[!htbp]
\centering
\includegraphics[height=5.0cm]{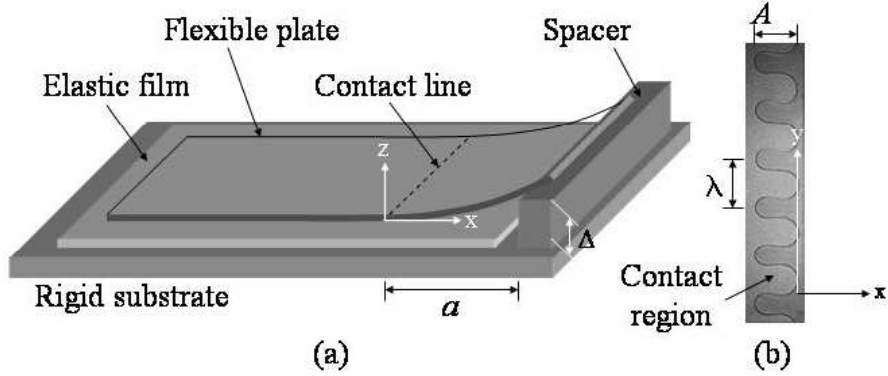}
\caption{(a) Schematic of the experiment in which a model elastic
adhesive remains bonded to a rigid substrate and a flexible plate
is detached from it with the help of a spacer inserted at the
opening of the crack. For a critically confined film, the contact
line does not remain smooth but becomes undulatory as shown in
video-micrograph (b).} \label{fig:fig1}
\end{figure}
The schematic of our experiment is represented in Figure
\ref{fig:fig1}a in which an elastic film of thickness $h$ and
shear modulus $\mu$ remains strongly bonded to a rigid substrate
while a flexible plate of rigidity $D$ in contact with the film in
the form of a curved elastica is supported at one end using a
spacer of height $\Delta$. The straight contact line between the
film and the plate becomes wavy when the film thickness $h$
decreases below a critical value $h_c$ or the curvature of the
plate decreases below a critical value $1/\rho_c$. Figure
\ref{fig:fig1}b represents a typical video-micrograph of such
undulations which are characterized by two different length
scales: the separation distance $\lambda$ between the waves which
scales as $\lambda \sim 4h$ and the amplitude $A$ which varies
with $D$ and $\mu$ as: $A \sim \left(D/\mu\right)^{1/3}$ (Ghatak
\textit{et al} 2003). The figure depicts also the co-ordinate
system in which $x$, $y$ and $z$ axes represent respectively the
direction of propagation of the contact line, the direction of the
wave vector and the thickness coordinate of the film. The $y$ axis
is located along the tips of the waves so that the film is
completely out of contact with the plate at $0<x<a$. Assuming the
film to be incompressible and purely elastic with no viscous
effect, we write the following stress equilibrium relations in the
absence of any body forces,
\begin{eqnarray}\label{eq:eq1}
  p_x & = & \mu \left( u_{xx} + u_{yy} + u_{zz} \right ) \nonumber \\
  p_y & = & \mu \left( v_{xx} + v_{yy} + v_{zz} \right ) \nonumber \\
  p_z & = & \mu \left( w_{xx} + w_{yy} + w_{zz} \right )
\end{eqnarray}
where, $u$, $v$ and $w$ are the displacements in the $x$, $y$ and
$z$ directions respectively and $p$ is the pressure. Here and
everywhere $s_x = \partial s/\partial x$ and $s_{xx} =
\partial^2 s /\partial x^2$. The incompressibility of the film results in:
\begin{eqnarray}\label{eq:eq2}
u_x + v_y + w_z = 0
\end{eqnarray}
These equations are solved using the following set of boundary
conditions (b.c.): (a) since the film remains strongly bonded to
the substrate we use no slip boundary condition at the interface
of the film and the substrate ($z=0$),
\begin{eqnarray}\label{bc.a}
u\left (z = 0\right) = v\left (z = 0\right) = w\left (z = 0\right)
=0
\end{eqnarray}
(b) we assume frictionless contact at the interface of the film
and the cover plate, $z = h$ which results in zero shear stress at
the interface,
\begin{eqnarray}\label{bc.b}
\sigma_{xz}\left(x,y,h\right) = 0 = \sigma_{yz}\left(x,y,h\right)
\end{eqnarray}
(c) we assume continuity of normal stress across the interface
($z= h$) which implies that the pressure is equal to the bending
stress on the plate
\begin{eqnarray}\label{bc.c}
p\left(z=h\right) = D \nabla^2 \psi\left(z=h\right) \hspace{4 mm}
\textrm{at} \hspace{2 mm} x<0
\end{eqnarray}
here $\nabla \equiv
\partial^2/\partial x^2 +\partial^2/\partial y^2 $ is the two
dimensional Laplacian, $D$ is the flexural rigidity of the plate
and $ \psi = \left. w\right |_{x,z=h}$ is its vertical
displacement. Since the plate bends only in the direction of $x$
co-ordinate its vertical displacement $\psi$ remains uniform along
the $y$ axis; hence, we simplify b.c. \ref{bc.c} as
\begin{eqnarray}\label{bc.c2}
p\left(z=h\right) = D \psi_{xxxx}\left(z=h\right) \hspace{4 mm}
\textrm{at} \hspace{2 mm} x<0
\end{eqnarray}
At $0<x<a$ there is no traction either on the film or the plate
which yields
\begin{eqnarray}\label{bc.d}
\left. \sigma_{xz}\right|_{z=h} = \left. \sigma_{yz}\right|_{z=h}
= \left. \sigma_{zz}\right|_{z=h} = 0
\end{eqnarray}
Equation (\ref{eq:eq1})-\ref{eq:eq2} can be written in
dimensionless form using the following dimensionless quantities,
\begin{eqnarray}
X=xq, \hspace{1 mm} Y = y/h,\hspace{1 mm} Z = z/h, \hspace{1 mm} U
= uq,\hspace{1 mm} V = v/h,\hspace{1 mm} W = w/h, \hspace{1 mm}
\Psi = \psi/h,\hspace{1 mm}P = p\epsilon^2/\mu \nonumber
\end{eqnarray}
While thickness $h$ of the film is the characteristic lengths
along the $y$ and $z$ axes, $q^{-1}$ is that along $x$. The length
$q^{-1}$ can be derived as the ratio of the deformability of the
plate and the film (Dillard \textit{et al} 1989, Ghatak \textit{et
al} 2004): $ q^{-1} = \left(Dh^3/3\mu\right)^{1/6} $. The quantity
$\epsilon = hq$ defined as the ratio of the two characteristic
lengths is a measure of the confinement of the film such that a
lower value of $\epsilon$ represents a more confined film.
Equations \ref{eq:eq1} and \ref{eq:eq2} can then be written in the
following dimensionless form
\begin{eqnarray}\label{eq:eq3}
  P_X & = & \epsilon^2 U_{XX} - U_{YY} + U_{ZZ} \nonumber \\
  P_Y & = & \epsilon^4 V_{XX} + \epsilon^2 \left( V_{YY} + V_{ZZ} \right ) \nonumber \\
  P_Z & = & \epsilon^4 W_{XX} + \epsilon^2 \left (W_{YY} + W_{ZZ} \right )
  \nonumber\\
  0 & = & U_X + V_Y + W_Z
\end{eqnarray}
while the boundary conditions \ref{bc.a}-\ref{bc.d} results
\begin{eqnarray}\label{bc.A}
\textrm{(a)} & U\left (Z = 0\right) = V\left (Z = 0\right) = 0 & \nonumber \\
\textrm{(b)} & \sigma_{XZ}\left(X,Y,Z=1\right) = 0 =
\sigma_{YZ}\left(X,Y,Z=1\right) & \nonumber\\
\textrm{(c)} & P\left(Z=1\right) = 3 \Psi_{XXXX} \hspace{4 mm}
\textrm{at} \hspace{2 mm} X<0 & \nonumber \\
\textrm{(d)} & 0 = \Psi_{XXXX} \hspace{4 mm} \textrm{at} \hspace{2
mm} 0<X<aq &
\end{eqnarray}
where $aq$ is the dimensionless crack length. Equation
\ref{eq:eq3} is solved by the regular perturbation technique which
assumes that the solutions consist of two components: the base
solutions which remain uniform along the $Y$ co-ordinate and the
correction term which incorporates the spatial variation along the
$Y$ axis. Thus the base solutions are of order $\epsilon^{0}$ and
the perturbed solutions are of order $\epsilon^2, \epsilon^4,
\ldots $, so that any variable $T\left(X,Y,Z\right) =
T_o\left(X,Z\right) + \epsilon^2 T_1\left(X,Y,Z\right) +
\epsilon^4 T_2\left(X,Y,Z\right)+ \ldots $ where $T = P$, $U$, $V$
and $W$. Inserting these definitions in equation \ref{eq:eq3} and
separating the base ($Y$ independent) and the perturbed ($Y$
dependent) terms yield:
\begin{eqnarray}\label{eq:eq4a}
{P_0}_X = \left( \epsilon^2 {U_0}_{XX} + {U_0}_{ZZ} \right
),\hspace{2mm} {P_0}_Z  = \left( \epsilon^4 {W_0}_{XX} +
\epsilon^2 {W_0}_{ZZ} \right ),\hspace{2mm} 0 = {U_0}_X + {W_0}_Z
\end{eqnarray}
and
\begin{eqnarray}\label{eq:eq4b}
 \epsilon^2 {P_1}_X + \epsilon^4 {P_2}_X & = &
 \epsilon^2 \left({U_1}_{YY} + {U_1}_{ZZ}\right) + \epsilon^4\left({U_1}_{XX} + {U_2}_{YY} + {U_2}_{ZZ}\right)
 + \epsilon^6 {U_2}_{XX}
 \nonumber \\
 \epsilon^2 {P_1}_Y + \epsilon^4 {P_2}_Y & = & \epsilon^4 \left( {V_1}_{YY} + {V_1}_{ZZ} \right ) +
 \epsilon^6 \left({V_1}_{XX} + {V_2}_{YY} + {V_2}_{ZZ}\right)\nonumber \\
 \epsilon^2 {P_1}_Z + \epsilon^4 {P_2}_Z & = & \epsilon^4 \left( {W_1}_{YY} + {W_1}_{ZZ} \right ) +
 \epsilon^6 \left({W_1}_{XX} + {W_2}_{YY} + {W_2}_{ZZ}\right) \nonumber \\
  0 & = & {U_1}_X + {V_1}_{Y} + {W_1}_Z
\end{eqnarray}
which are solved using b.c. derived from equations \ref{bc.A}a-d.

\textbf{Base Solution:} Since for a thin film, $\epsilon^2 << 1$,
equation \ref{eq:eq4a} can be simplified by neglecting the terms
containing $\epsilon^2$. Integration of the resulting equations
(presented in detail in reference Ghatak \textit{et al} 2004)
finally leads to the following solution for the base components of
the displacements in the film and the plate,
\begin{eqnarray}\label{eq:eq12a}
U_{0} & = &\left(3Z^2/2 - 3Z\right)F' \phi_{1} \left(X\right),
\hspace{5mm} W_{0} = \left(3Z^2/2 - Z^3/2\right)F' \phi_{2}
\left(X\right)\nonumber \\
\phi_1 & = & e^{X/2}\left(aq e^{X/2} +
\left(3aq+4\right)\sin\left(\sqrt{3}X/2\right)/\sqrt{3} -aq
\cos\left(\sqrt{3}X/2\right)\right) \nonumber \\
\phi_2 & = & e^{X/2}\left(aq e^{X/2} +
\left(3aq+2\right)\sin\left(\sqrt{3}X/2\right)/\sqrt{3} +
\left(aq+2\right) \cos\left(\sqrt{3}X/2\right)\right)
\nonumber \\
F' & = & 3\bar{\Delta}/\left(6 + 12 aq + 9\left(aq\right)^2 +
2\left(aq\right)^3\right) \nonumber \\
\Psi_{0} & = & F' \phi_2 \left(X\right) \hspace{2 mm} \textrm{$X < 0$} \nonumber \\
         & = & F'\left(2\left(aq+1\right) + \left(3aq+2\right)X +
aqX^2 - X^3/3 \right) \hspace{2 mm} \textrm{$0<X<aq$}
\end{eqnarray}
which suggest oscillatory variation along $X$ with exponentially
vanishing amplitude away from the contact line. Such displacements
results in the following expression for the dimensionless work of
adhesion (Ghatak \textit{et al} 2004) $G =
W_A/\left(\mu/q\right)$:
\begin{eqnarray}\label{eq:eq13}
G & = & g\left(aq\right)\left(27
\bar{\Delta}^2/2\epsilon\left(aq\right)^4\right)
\nonumber \\
g\left(aq\right ) & = & 8\left(aq\right)^4\left(12 + 46
\left(aq\right) + 72 \left(aq\right)^2 + 56\left(aq\right)^3 +
21\left(aq\right)^4+3\left(aq\right)^5\right)/\nonumber
\\
& & 3\left(6+12\left(aq\right)+9\left(aq\right)^2 +
2\left(aq\right)^3\right)^3
\end{eqnarray}
in which $g\left(aq\right)$ is the correction to the classical
result of Obreimoff (Obreimoff \emph{et al} 1930) for peeling off
a rigid substrate.

\textbf{Perturbation Analysis:} Matching the coefficients for
$\epsilon^{i}$, $i = 2, 4$ in the left and the right hand side of
equation \ref{eq:eq4b} results in the following set of equations:
\begin{equation}\label{eq:eq22}
\begin{array}{llll}
\epsilon^2: & {P_1}_X = {U_1}_{YY} + {U_1}_{ZZ}, & {P_1}_Y = 0,
& {P_1}_Z = 0 \\
\epsilon^4: & \hspace{-1mm}{P_2}_{X} = {U_1}_{XX} + {U_2}_{YY} +
{U_2}_{ZZ}, & \hspace{-1mm}
{P_2}_{Y} = {V_1}_{YY} + {V_1}_{ZZ}, & \hspace{-1mm} {P_2}_{Z} = {W_1}_{YY} + {W_1}_{ZZ}\\
& {U_1}_{X} + {V_1}_{Y} + {W_1}_{Z} = 0 & &
\end{array}
\end{equation}
we assume also that the excess quantities vary sinusoidally along
the $Y$ axis: so that $T_i = \bar{T_i} \sin\left(KY\right)$; $T =
U$, $V$, $W$, $P$; $i = 1$, $2$; $K = 2\pi/\left(\lambda/h\right)$
is the dimensionless wave number of the perturbed waves. Equations
\ref{eq:eq22} are solved using the following boundary conditions
derived from \ref{bc.A},
\begin{eqnarray}\label{eq:eq21}
\begin{array}{llll}
\textrm{(a)}& \textrm{at} \hspace{2 mm} Z = 0 & \hspace{2mm}
\bar{U}_1 = \bar{V}_1 = \bar{W}_1 = \bar{U}_2 = 0 & \\
\textrm{(b)}& \textrm{at} \hspace{2 mm} Z = 1 \hspace{2mm}
{\bar{U_1}}_Z  = 0 & \left({\bar{U_2}}_Z +
{\bar{W_1}}_X\right) = 0 \hspace{3 mm} \left({\bar{V_1}}_Z + {\bar{W_1}}_Y\right) = 0 & \\
\textrm{(c)}& \textrm{at} \hspace{2 mm} Z = 1, X<0 & \hspace{-2
mm} \bar{P}_1 = 3 {\Psi_1}_{XXXX}, & \hspace{-32 mm} \textrm{at}
\hspace{2 mm} 0<X<aq \hspace{2 mm} 0 = {\Psi_1}_{XXXX}
\end{array}
\end{eqnarray}
where $\Psi_{1} = \bar{W_1}\left(X,Z=1\right)$ is the vertical
displacement of the flexible plate at $Z=1$. Assumption of
sinusoidal dependence on $Y$ is a simplification which is apparent
from the video micrographs of the contact line as in figure
\ref{fig:fig1}b which show that the film and the plate remains in
contact whole through the area of the finger implying that the
deformation of the film is not perfectly sinusoidal, as it would
mean a line contact between the plate and the film. However, here
I assume sinusoidal variation to keep the calculations simple.
Since, the plate does not bend in the direction of $Y$, $\Psi_{1}$
and ${\Psi_1}_{XXXX}$ both remain uniform along this axis;
consequently in b.c. \ref{eq:eq21}c, bending stress on the plate
is equated to $\bar{P}_1$. Equation \ref{eq:eq22} suggests that
$P_{1}$ remains independent of $Y$ and $Z$ although $U_1$ varies
along $Y$, hence, the only solution for $P_{1}$ that can satisfy
equation \ref{eq:eq22} is $P_{1} = 0$. Other components of the
excess displacements are obtained as:
\begin{eqnarray}\label{eq:eq24}
U_1 & = & 0  \nonumber \\
V_1 & = & \frac{C\left(X\right)}{K} \left( \left(\frac{-2K
\left(e^{K} + e^{-K}\right)}{e^{K} + e^{-K} + 2K e^{K}} \right)
\sinh\left(KZ\right) + \right. \nonumber \\
  &   & \left. KZ \left( \frac{e^{K} + e^{-K} - 2K e^{-K}}{e^{K} +
e^{-K} + 2Ke^{K}} e^{KZ} - e^{-KZ} \right) \right) \cos\left(KY\right) \nonumber \\
W_1 & = & \frac{C\left(X\right)}{K} \left( -\frac{2e^{K} + 2e^{-K}
+ 2K\left(e^{K} - e^{-K} \right)}{e^{K} + e^{-K} + 2Ke^{K}}
\sinh\left(KZ\right) + \right. \nonumber \\
  &   & \left. KZ \left( \frac{e^{K} + e^{-K} - 2K e^{-K}}{e^{K} +
e^{-K} + 2K e^{K}} e^{KZ} + e^{-KZ}\right)
\right)\sin\left(KY\right)
\end{eqnarray}
Here, only the lowest order terms in $\epsilon$ are computed since
the higher order terms enhance accuracy insignificantly. Since for
all our experiments, $\epsilon < 0.3$, the above solutions imply
that the excess deformations in the film occur under very small
excess pressure which is of the order $\epsilon^4 < 0.01$. This
excess pressure, however small, varies along $Y$ implying that it
should depend upon the distance between the plate and the film.
Nevertheless, the excess traction which results from the distance
dependent forces (Shenoy et al 2001, Sarkar et al 2004) apply only
in the immediate vicinity ($<0.1 \mu$m) of the contact between the
film and the plate as the gap between the two increases rather
sharply (could be observed in AFM images of the permanent patterns
of surface undulations). Hence, it does not contribute any
significantly to the overall energetics.

While equation \ref{eq:eq24} elaborates the variation of excess
deformations along $Y$ and $Z$, their dependence on $X$ is
incorporated through the coefficient $C\left(X\right)$ which is
obtained by solving equations \ref{eq:eq21}c using the following
boundary conditions
\begin{equation}\label{eq:eq25}
\begin{array}{llll}
\textrm{(i, ii)} \left. \Psi_1\right |_{X=0} = C_0
\Phi\left(K\right) & \textrm{(iii)} \left.
{\Psi_1}_{X}\right|_{X=0-} = \left. {\Psi_1}_{X}\right|_{X=0+} &
\nonumber \\ \textrm{(iv)} \left. {\Psi_1}_{XX}\right|_{X=0-} =
\left. {\Psi_1}_{XX}\right|_{X=0+} & \textrm{(v, vi)} \left.
\Psi_1\right|_{X=-\xi} = \left. {\Psi_1}_X\right|_{X=-\xi} = 0 &
\nonumber \\
\textrm{(vii, viii)} \left.\Psi_1\right|_{X=aq} =
\left.{\Psi_1}_{XX}\right|_{X=aq} = 0 & &
\end{array}
\end{equation}
Where, $\xi = Aq$ is the dimensionless amplitude of the waves and
$C_0 \Phi\left(K\right) $ is the excess stretching of the film at
$x=0$: $C_0$ is a constant and $\Phi\left(K\right) = \left.
\bar{W}_1/C\left(X\right)\right|_{X=0} =
\left(e^{-K}-e^{3K}+4Ke^{K}\right)/K\left(1+e^{2K}\left(1+2K\right)\right)$.
(a) B.C. i, ii, iii and iv occur because the excess displacement
and slope of the plate are continuous at $X=0$; (b) at $X=-\xi$
the displacement of the plate, its slope should vanish, which
result in the b.c. v and vi; (c) similarly, at $X = aq$ excess
displacement and curvature of plate is zero, which results in b.c.
vii and viii. Incorporating these boundary conditions into the
solutions of equations \ref{eq:eq21}c yield the following
expression for the the excess displacement of the plate:
\begin{eqnarray}\label{eq:eq27}
\Psi_1  & = & \frac{C_0 \Phi\left(K\right)}{aq \left(3\xi +
4aq\right)} \left(-\left(3\xi^2+ 6aq \xi + 2
\left(aq\right)^2\right)\left(X/\xi\right)^3 - 3\left(2\xi^2+3aq
\xi\right)\left(X/\xi\right)^2 \right.
\nonumber \\
& & \left. - 3\left(\xi^2 - 2
\left(aq\right)^2\right)\left(X/\xi\right) + aq \left(3 \xi + 4 aq
\right)\right) \hspace{2mm} \textrm{at} \hspace{2mm} X<0 \nonumber \\
\Psi_1  & = & C_0 \Phi\left(K\right) \left(1 +
\frac{aq}{\xi\left(3\xi + 4aq\right)} \left(\left(2\xi +
3aq\right) \left(X/aq\right)^3 - \right. \right. \nonumber \\
& & \hspace{-10mm} \left. \left. 3\left(2\xi + 3aq\right)
\left(X/aq\right)^2 - 3\left(\xi^2 - 2
\left(aq\right)^2\right)\left(X/\left(aq\right)^2\right) \right)
\right) \hspace{2mm} \textrm{at} \hspace{2mm} 0<X<aq
\end{eqnarray}

\textbf{Excess Energy:} Total energy of the system consists of the
elastic energy of the film, bending energy of the plate and the
interfacial energy:
\begin{eqnarray}\label{eq:eq41}
\Pi & = & \Pi_{e} + \Pi_{b} + \Pi_{i} \nonumber \\
   & = & \frac{\mu}{4}
{\int_{-\infty}}^{0} {\int_{0}}^{2\pi/k}{\int_{0}}^{h} \left(
\left(v_z + w_y \right)^{2} + \left(u_y + v_x\right)^{2} + \left(
u_z +
w_x\right)^{2} \right ) dz dy dx \nonumber \\
& & + \frac{D}{2} {\int_{-\infty}}^{a} {\int_{0}}^{2\pi/k} \left(
{\psi}_{xx}\right)^2 dy dx + W_A\left(2\pi a/k +
A_{\textrm{finger}}\right)
\end{eqnarray}
where $A_{\textrm{finger}}$ is the interfacial area of contact at
$-a<x<0$. From equation \ref{eq:eq41} the excess energy of the
system is obtained as $\Pi_{excess} = \Pi - \Pi_{0}$ which is
written in a dimensionless form using $\mu/q^3$ as the
characteristic energy and by substituting for variables $T = T_0 +
\epsilon^2 T_1 + \epsilon^4 T_2$, where $T = U$, $V$, $W$ and $P$.

The expression for excess elastic energy $\bf{\Pi_{e}}$ in the
film is obtained as,
\begin{eqnarray}\label{eq:eq42}
\bf{\Pi_e} & = & \frac{1}{4} {\int_{-\xi}}^{0}
{\int_{0}}^{2\pi/K}{\int_{0}}^{1} \left( \epsilon^6  \left({V_1}_Z
+ {W_1}_Y + \epsilon^2\left({V_2}_Z + {W_2}_Y\right) \right)^{2} +
\right. \nonumber \\
& & \left. \epsilon^8 \left( {U_2}_Y + {V_1}_X + \epsilon^2
{V_2}_X\right)^{2} + \epsilon^8 \left( \left({U_2}_Z +
{W_1}_X\right) + \epsilon^2 {W_2}_X\right)^{2} + \right. \nonumber
\\
& & \left. 2
\epsilon^4\left({U_0}_Z+\epsilon^2{W_0}_X\right)\left(
{U_2}_Z+{W_1}_X + \epsilon^2 {W_2}_X\right) \right ) dZ dY dX
\end{eqnarray}
where I
 estimate the excess energy within a distance $-\xi \leq X
\leq 0$. Considering only the leading order terms ($\epsilon^4$
and $\epsilon^6$) I can simplify the expression in \ref{eq:eq42}
as,
\begin{eqnarray}\label{eq:eq43a}
\bf{\Pi_e} & = & \frac{\epsilon^6}{4} {\int_{-\infty}}^{0}
{\int_{0}}^{2\pi/K}{\int_{0}}^{1} \left({V_1}_Z + {W_1}_Y
\right)^{2} dZ dY dX
\end{eqnarray}
in which I substitute the expressions for $V_1$ and $W_1$ from
\ref{eq:eq24} to obtain
\begin{eqnarray}\label{eq:eq43b}
\bf{\Pi_e} & = & \epsilon^6 {C_0}^2 f_2\left(\xi,aq,K\right)/4 =
\epsilon^6 {C_0}^2 f_0\left(K\right)
\bar{f_2}\left(\xi,aq\right)/4
\end{eqnarray}

Similarly, the dimensionless excess bending energy ${\bf \Pi_b}$
of the plate:
\begin{eqnarray}\label{eq:eq44}
{\bf \Pi_b} = \left(3\pi/K\right) {\int_{-xi}}^{aq}
\left(2\epsilon^2{\Psi_0}_{XX}{\Psi_1}_{XX} +
\epsilon^4\left({\Psi_1}_{XX}+ \epsilon^2
{\Psi_2}_{XX}\right)^2\right) dX
\end{eqnarray}
Considering only the leading order terms: $\epsilon^2$ and
$\epsilon^4$, equation \ref{eq:eq44} simplifies to
\begin{eqnarray}\label{eq:eq45}
{\bf \Pi_b} = \frac{3\pi}{K} {\int_{-xi}}^{aq}
\left(2\epsilon^2{\Psi_0}_{XX} {\Psi_1}_{XX} + \epsilon^4
{\Psi_1}_{XX}^2\right) dX
\end{eqnarray}
Substituting the expressions for $\Psi_0$ and $\Psi_1$ from
\ref{eq:eq12a} and \ref{eq:eq27}, ${\bf \Pi_b}$ is obtained as,
\begin{eqnarray}\label{eq:eq45b}
{\bf \Pi_b} = \epsilon^2 C_0 \bar{\Delta} f_3\left(\xi, aq, K
\right) + \epsilon^4 {C_0}^2 f_1\left(\xi, aq, K \right)
\end{eqnarray}

The interfacial energy is estimated as,
\begin{eqnarray}\label{eq:eq46}
{\bf \Pi_i}  = 2G\epsilon \int_0^{\pi/K}\frac{\xi}{2}
\sin\left(KY\right) dY= \frac{2G\xi\epsilon}{K}
\end{eqnarray}

Substituting the expression for $G$ from equation \ref{eq:eq13} in
equation \ref{eq:eq46} and combining all the three energies,
yields the total excess energy as:
\begin{eqnarray}\label{eq:eq47}
{\bf \Pi} = \left(\epsilon^6 f_2\left(\xi, aq, K\right) +
\epsilon^4 f_1\left(\xi, aq, K \right) \right) {C_0}^2 +
\epsilon^2 f_3\left(\xi, aq, K \right) \bar{\Delta} C_0  +
\frac{27 \bar{\Delta}^2 \xi
}{K}\frac{g\left(aq\right)}{\left(aq\right)^4}
\end{eqnarray}
The expressions for $f_1\left(\xi,aq,K \right)$,
$f_2\left(\xi,aq,K\right)$ and $f_3\left(\xi,aq,K \right)$ are
obtained using Mathematica and are not being presented here since
they could not be written in a compact form.

\section{Results and Discussion}
The expression for excess energy in equation \ref{eq:eq47}
accounts for the combined effects of three sets of parameters:
$\epsilon$, the confinement parameter, $\xi$ and $K$, the
characteristic length scale of the perturbations and $aq$ and
$\bar{\Delta}$, the length scale of the geometry of the
experiment. In what follows, we look for the solutions of these
different sets of parameters which results in negative excess
energy associated with the instability.

While it is evident from equation \ref{eq:eq47} that $\xi$ and $K$
are nonlinearly coupled quantities, the physics of the problem is
better understood if we study their effects separately. We do that
following the observation that the excess displacements in the
film but not that of the plate are functions of $Y$, which allows
us to assume that the dimensionless wave number $K$ of the
perturbations is determined solely by the minima of the excess
elastic energy ${\bf \Pi_{e}}$ and not the other components of the
total excess energy. Although this assumption is not exactly
correct as the displacement $\Psi$ of the plate is also a function
of $K$, experimental observation that the amplitude remains nearly
independent of the wavelength (Ghatak \textit{et al} 2003)
suggests that the above assumption should not insert much
inaccuracy into the calculation.

\begin{figure}[!htbp]
\centering
\includegraphics[width=6.1cm]{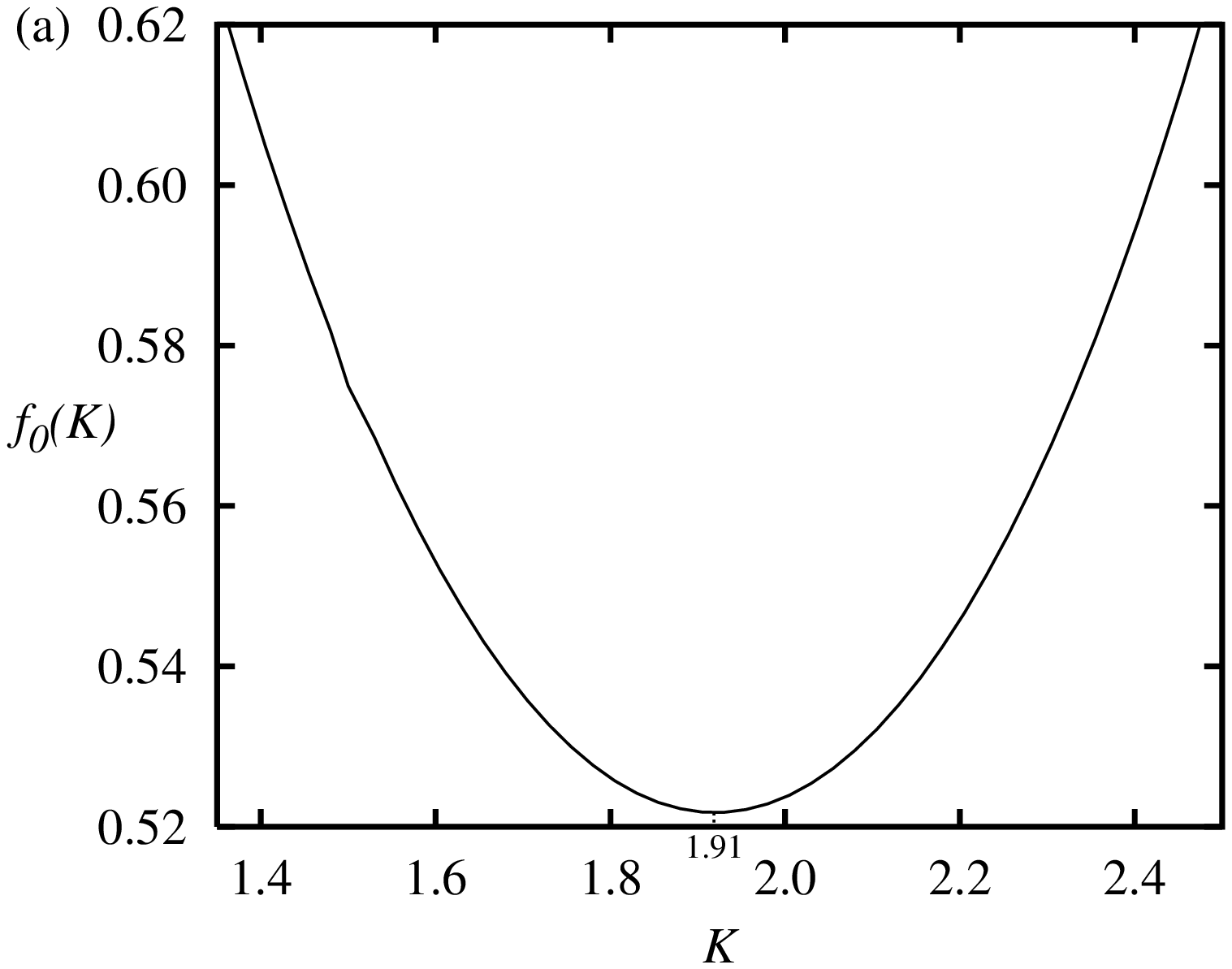}
\includegraphics[width=6.2cm]{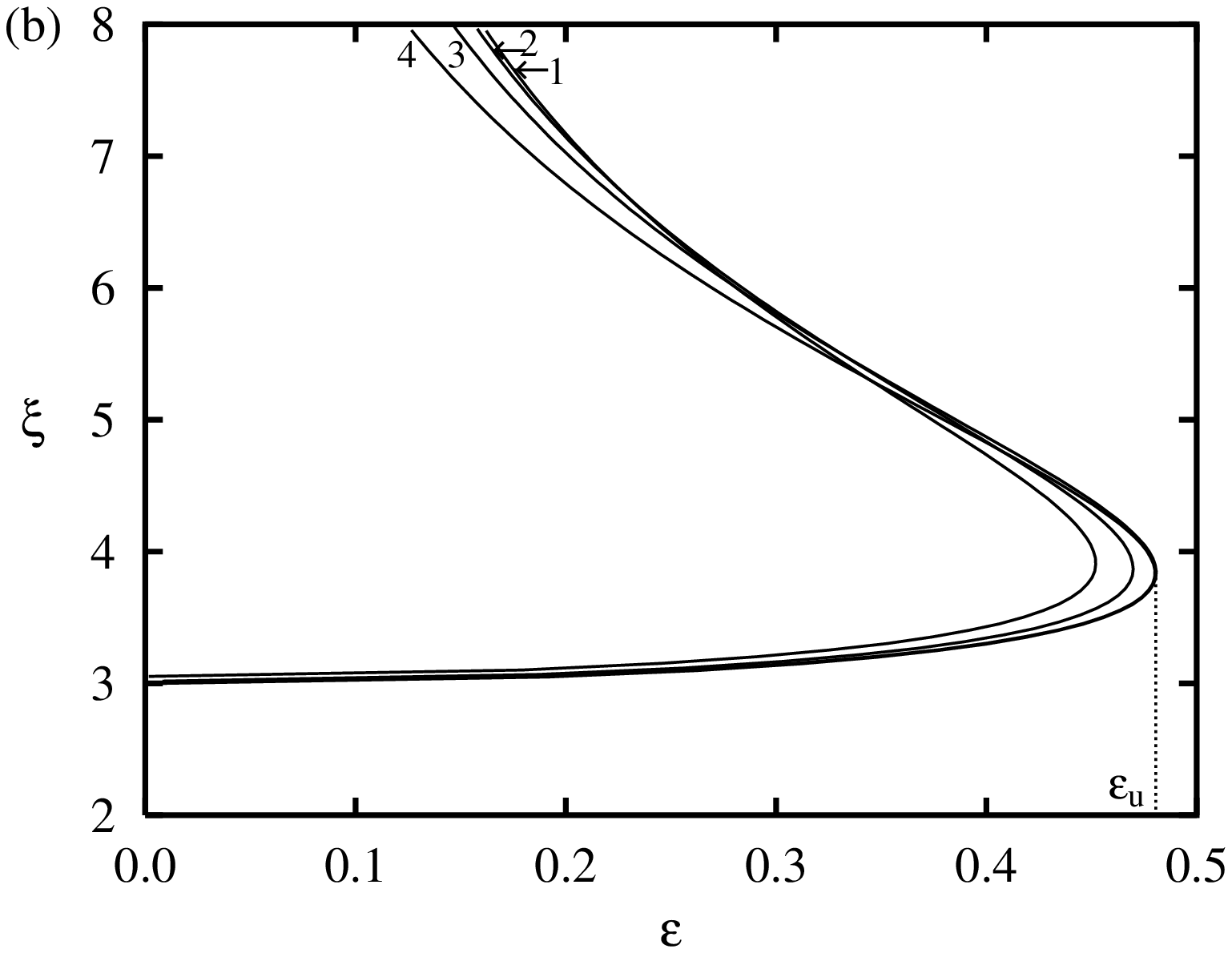}
\caption{(a) Dimensionless elastic energy of the film is plotted
against the dimensionless wave number $K$ of the surface
undulations. The energy of the film attains a minima at $K=1.91$
implying that the wavelength varies with thickness of the film as
$\lambda = 3.3h$. (b)Amplitude $\xi$ are plotted against the
limiting values of $\epsilon$ from equation \ref{eq:eq49} for
different values of the dimensionless length $aq$. Curves 1-4
represent $aq = 5$, $15$, $25$ and $55$ respectively. $\epsilon_u
= 0.48$ is the upper limit for $\epsilon$ beyond which no real
solution for $C_0$ exists.} \label{fig:fig4}
\end{figure}
The plot of excess elastic energy ${\bf \Pi_{e}}$ vs. the wave
number $K$ in figure \ref{fig:fig4}a then shows that $\bf{\Pi_e}$
attains a minima when $K =1.91$. The wavelength of perturbations
thus scales with the thickness of the film as $\lambda = 3.3 h$
which corroborates with the general observation in a wide range of
experimental geometries (Ghatak \textit{et al} 2000, 2003,
M$\ddot{o}$nch \textit{et al} 2001) that $\lambda$ remains
independent of all the material and the geometric properties of
the system except $h$. Furthermore, the proportionality constant
matches well with that observed in experiments ($\lambda =
3-4h$)with rigid and flexible contacting plates.

Although the minima of ${\bf \Pi_{e}}$ occurs at $K=1.91$
irrespective of $C_{0}$ and $\epsilon$, for both these parameters
real values are desired. In fact, equation \ref{eq:eq47}, being
quadratic w.r.t. $C_{0}$, suggests that in the limit ${\bf \Pi} =
0$ the real solutions for $C_{0}$ exist only when,
\begin{eqnarray}\label{eq:eq48}
\left(\epsilon^2 f_3\left(\xi, aq, K \right)\right)^2 - 4
\left(\epsilon^6 f_2\left(\xi, aq, K\right) + \epsilon^4
f_1\left(\xi, aq, K \right)\right)\left(27 \xi/K\right)
g\left(aq\right)/\left(aq\right)^4 \geq 0
\end{eqnarray}
resulting in the following inequality for $\epsilon^2$:
\begin{eqnarray}\label{eq:eq49}
\epsilon^2 \leq \frac{{f_3}^2\left(\xi, aq, K
\right)}{4f_2\left(\xi, aq, K\right)}
\frac{K}{27\xi}\frac{\left(aq\right)^4}{g\left(aq\right)} -
\frac{f_1\left(\xi, aq, K \right)}{f_2\left(\xi, aq, K\right)}
\end{eqnarray}
Equation \ref{eq:eq49} sets an upper bound for $\epsilon$ as
evident from figure \ref{fig:fig4}b where $\xi$ and $\epsilon$
which satisfy equation \ref{eq:eq49} are plotted for $K=1.91$ and
for different $aq$. When $\epsilon$ is smaller than this upper
critical limit $\epsilon_u$, two different solutions for $\xi$
exist, the stability of which depends upon whether ${\bf \Pi}$
attains a minima at these solutions. Hypothesizing that ${\bf
\Pi}$ minimizes when $\partial {\bf \Pi}/\partial C_0 = 0$, we
obtain an expression for $C_0$ which when substituted in
\ref{eq:eq47}, yields
\begin{eqnarray}\label{eq:eq51}
{\bf \Pi} = - \frac{\bar{\Delta}^2}{4}\frac{{f_3\left(\xi, aq, K
\right)}^2}{f_1\left(\xi, aq, K \right) + \epsilon^2 f_2\left(K,
\xi\right)} + \frac{27 \bar{\Delta}^2 \xi }{K}g\left(aq\right)
\end{eqnarray}
\begin{figure}[!htbp]
\centering
\includegraphics[width=6.0cm]{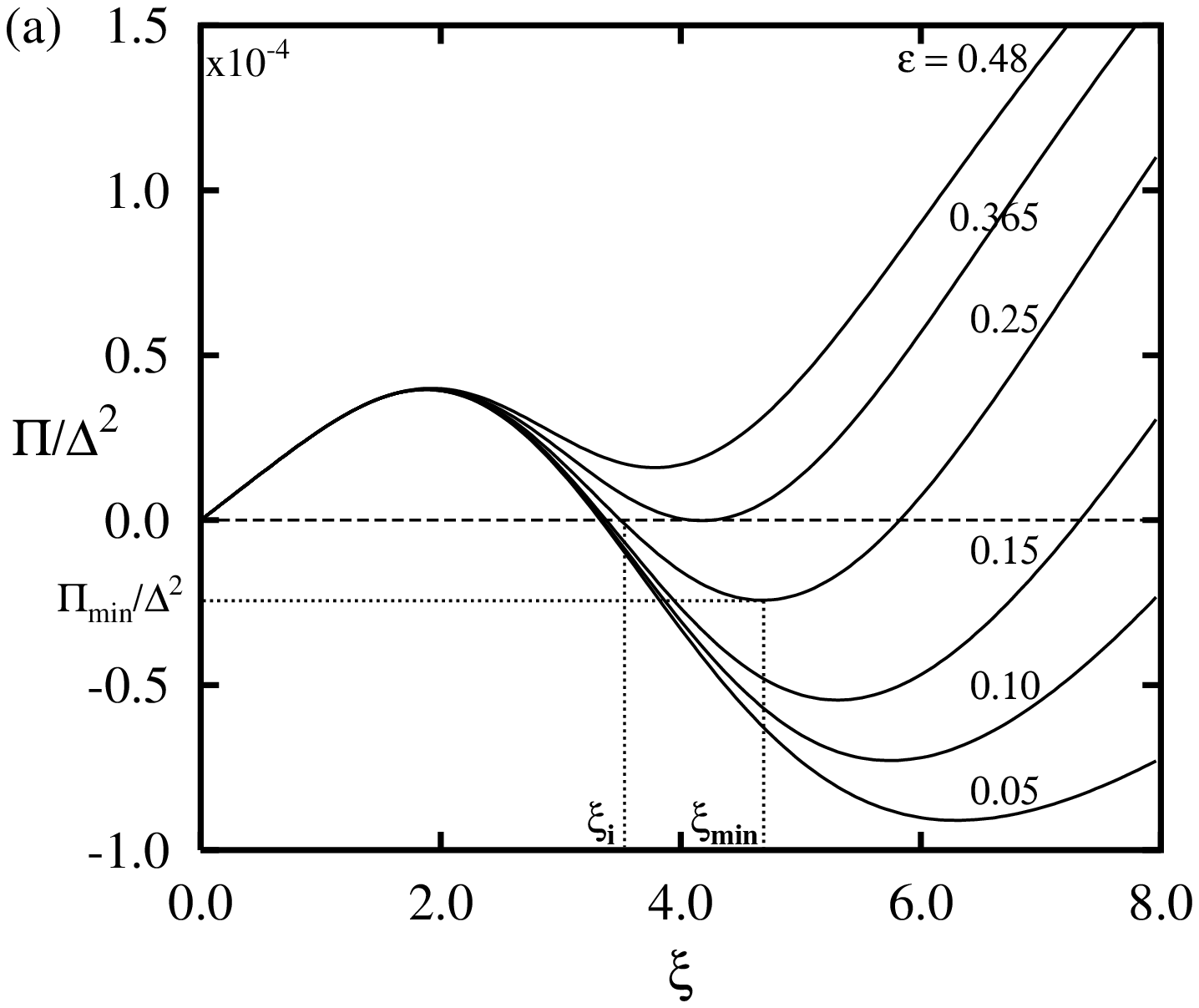}
\includegraphics[width=6.5cm]{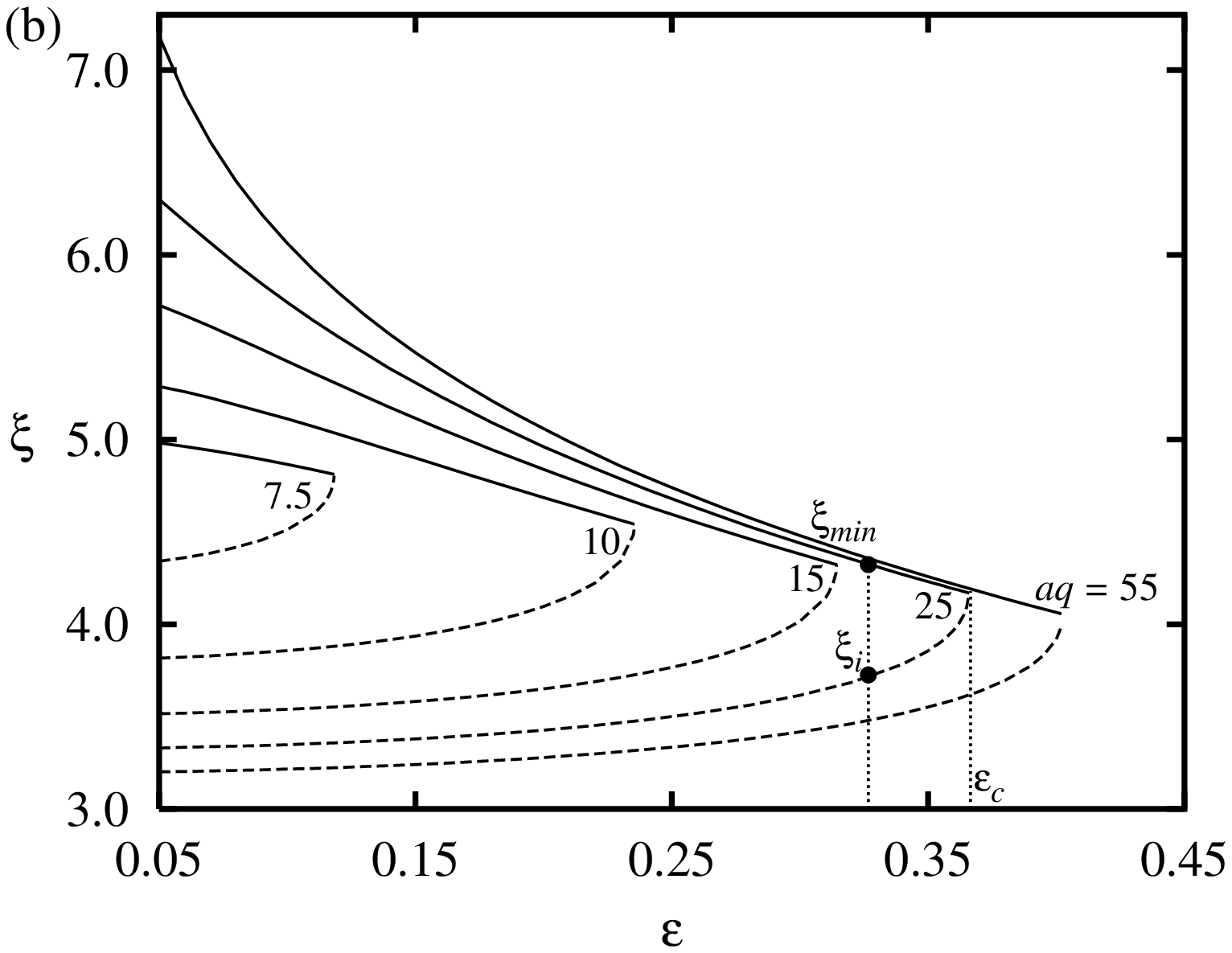}
\caption{(a) Dimensionless excess energy is plotted against the
dimensionless amplitude of the waves for different values of the
confinement parameter $\epsilon$. The curves are obtained using
representative values for the dimensionless parameters: $aq = 25$,
and $K=1.91$. (b) Bifurcation diagram showing variation of $\xi$
w.r.t. $\epsilon$ for different values of $aq$ and $K = 1.91$. The
dotted and the solid lines represent respectively $\xi_{i}$ vs.
$\epsilon$ and $\xi_{min}$ vs. $\epsilon$. No solution for $\xi$
exists beyond $\epsilon_{c}$.} \label{fig:fig5}
\end{figure}
In figure \ref{fig:fig5}a we plot ${\bf \Pi}$ from equation
\ref{eq:eq51} w.r.t. $\xi$, for $\epsilon = 0.48-0.05$. For all
these cases, ${\bf \Pi}$ exhibits a non-monotonic character: with
increase in $\xi$, it first increases till it reaches a maxima
after which it decreases to attain a minima, there after it
increases again. Stability of these systems is determined by the
minima of the excess energy which if positive signifies stable
base solution for the contact line and unstable solution if
negative. For example, for $\epsilon = 0.48$ ${\bf \Pi}$ remains
positive all-through implying that the undulation of a straight
contact line would increase the total energy of the system so that
a straight contact line would remain stable. On the other hand,
$\epsilon = 0.365$, presents a limiting case for which minima of
the excess energy ${\bf \Pi_{min}}$ attains zero and for $\epsilon
= 0.25$, $0.15$, $0.1$ and $0.05$, ${\bf \Pi_{min}}$ becomes
negative, i.e. when the film is more than critically confined i.e.
$\epsilon <\epsilon_{c} = 0.365$, the contact line can become
unstable if sufficiently perturbed. The critical value
$\epsilon_{c} = 0.365$ thus obtained for the above set of data
corroborates well with 0.3 obtained in experiments of figure
\ref{fig:fig1}. Furthermore, a finite energy barrier at $\xi <
\xi_{i}$ suggests that a straight contact line is not unstable for
perturbations of all magnitude. Because, for perturbation with
amplitude $\xi < \xi_{i}$ the excess energy remains positive so
that these perturbations decay to zero. This result too
corroborates with experiment that with increase in confinement of
the film, the amplitude of the undulations never increases from
exactly zero, but from a finite value. When $\xi > \xi_{i}$, ${\bf
\Pi}$ decreases to become negative, so that these perturbations
can grow till $\xi$ reaches $\xi_{min}$, at which ${\bf \Pi}$
attains the minima ${\bf \Pi_{min}}$; $\xi_{min}$ is then the
predicted amplitude of the undulations of the contact line.

Figure \ref{fig:fig5}b depicts the bifurcation diagram where
$\xi_{i}$ (dashed line) and $\xi_{min}$ (solid line) are plotted
with respect to $\epsilon$ for variety of $aq$. The dashed lines
signify that the perturbations whose amplitude $\xi < \xi_{i}$
decay to zero, whereas solid lines mean those with $\xi > \xi_{i}$
grow to $\xi_{min}$. The amplitude $\xi_{min}$ increases with
increase in the confinement of the film similar to that observed
in experiments, although, the values predicted are somewhat (~2-3
times) larger than what is observed. This discrepancy could be due
to the underestimation of the excess elastic energy of the film.
\begin{figure}[!htbp]
\centering
\includegraphics[height=5.5cm]{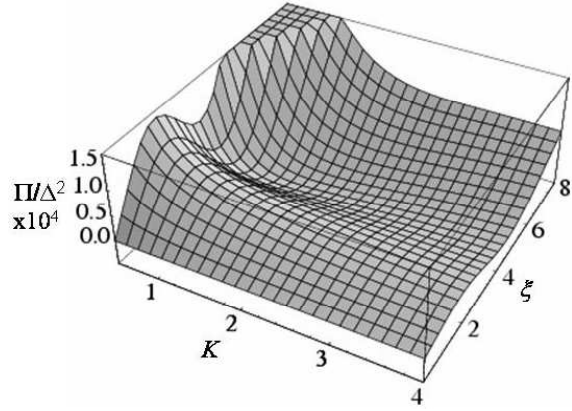}
\caption{Excess energy ${\bf \Pi}$ for $\epsilon=0.2$ and $aq =
25$ is plotted w.r.t amplitude $\xi$ and wave number $K$.}
\label{fig:fig8}
\end{figure}
In figure \ref{fig:fig8} the combined effects of $\xi$ and $K$ are
probed by plotting ${\bf \Pi}$ w.r.t. $\xi$ and $K$. Here again
${\bf \Pi_{min}}$ becomes negative for confinement parameter
$\epsilon$ below a critical value $\epsilon_c = 0.365$. However,
the wave number $K$ at which the minima occurs does not remain
constant, it decreases from 2.12 to 0.5 while $\epsilon$ varies
from 0.365 to 0.05. Although this prediction is somewhat different
from experiments, in which $\epsilon$ varies between $0.3$ to
$0.1$ at which $K$ is observed to be $1.57\pm 0.1$, some recent
observations (personal communication with Prof. A. Sharma) with
very thin elastic films ($\sim 0.5 \mu$m) indeed indicate that $K$
can decrease to $1.0$ as $\epsilon$ decreases to $0.07$. More
experiments are clearly necessary to characterize quantitatively
the effect of the coupling of the two length scales with highly
confined elastic films.

\section{Summary}
The analysis shows that confinement of an incompressible elastic
film leads to favorable energetics for perturbations to grow so
that the film can not deform uniformly everywhere when subjected
to the tensile stresses at the interface. Furthermore, the nature
of the adhesion stress is not important, even the spatial
variation of the surface forces play rather an insignificant role.
While the theory captures the essential physics of the problem,
slight overestimation of the amplitude possibly results from the
assumption of sinusoidal variations along the $Y$ axis which is
not perfectly correct. These issues can possibly be resolved by 3D
simulation of the force field near the contact line. Nevertheless
the results presented in this paper should be important for many
other systems in confined geometries.

\textbf{Acknowledgement:} I sincerely thank Prof. M. K. Chaudhury
in whose laboratory at Lehigh University and under whose guidance
all experiments were carried out. I thank also Prof. L. Mahadevan
for suggesting the perturbation analysis for solving the
elasticity equations. Many thanks to Prof. Asutosh Sharma and
Prof. V. Shankar for many stimulating discussions.
\end{document}